# Optimal Crosstalk Detection and Localization Method for Optical Time Division Multiplexed Transmission Systems

A. Ahmed JEDIDI and B. Mohamed ABID

**Abstract**— All-Optical Network (AON) is a network where the user-network interface is optical and the data does undergo optical to electrical conversion within the network. AONs are attractive because they promise very high rates, flexible switching and broad application support. There are two technologies for AON: Wavelength Division Multiplexed (WDM) and Optical Time Division Multiplexed (OTDM). OTDM transmission systems are becoming increasingly important as one of the key technologies satisfying the growing demand for large capacity optical networks. Although OTDM has several advantages in terms of operation system, such as natural accommodation of higher bit rate payloads, it introduces many security vulnerabilities, which do not exist in traditional networks. One of the serious problems with OTDM is the fact that optical crosstalk is additive, and thus the aggregate effect of crosstalk over a whole all-optical network (AON) may be more nefarious than a single point of crosstalk. This is because crosstalk can spread rapidly through the network, causing additional awkward failures and triggering multiple undesirable alarms. This results in the continuous monitoring and identification of the impairments becoming challenging in the event of transmission failures. In this paper we propose a novel approach for detecting and localizing crosstalk in OTDM transmission systems that can participate in some tasks for fault management in optical network.

---

## 1 INTRODUCTION

All-Optical Networks are emerging as a promising technology for very high data rate communications, flexible switching and broadband application support. They contain only AON components and therefore largely different from current optical networks. In particular, AONs provide transparency[1] capabilities and new features allowing routing and switching of traffic without any examination or modification of signals within the network. Transparency in AONs is desirable in many respects. It is specifically very useful for heterogeneous network environments to share network resources [1]. A lightpath[2], for example, will be capable of carrying different types of signals with arbitrary bit rates and protocol formats. Thus, terminal systems may employ different modulation formats and hence be upgraded without additional cost and complexity.

AONs have unique features and requirements in terms of security and management that distinguish them from traditional optical networks. In particular, the unique characteristics and behaviours of AON components bring an additional set of new challenges in network security. By their nature, AON components are vulnerable to various forms of service disruption, Quality of Service (QoS) degradation, and eavesdropping attacks [2]-[3]. Although network transparency offers many advantages for high data rate communications, it introduces many security vulnerabilities and miscellaneous transmission impairments such as optical crosstalk, Amplified Spontaneous Emission (ASE) noise and gain competition [4]. As a result, these impairments aggregate at several intermediate nodes and can impact the signal quality enough to reduce the QoS without preventing all network services.

An important implication of using AON components in optical communication systems is that available methods used to manage and monitor the health of the network may no longer be appropriate. One of the serious problems with transparency is the fact that optical crosstalk is additive, and thus the aggregate effect of crosstalk over a whole AON may be more nefarious than a single point of crosstalk [2]-[5]. Therefore, efficient monitoring and estimation of signal quality along a lightpath are of highest interest because of their importance in diagnosing and assessing the overall health of the network.

[1] A component is called X-transparent if it forwards incoming signals from input to output without examining the X aspect of the signal. For example, AON components are electrically transparent.

[2] A lightpath is defined as an *end-to-end* optical connection between a source and a destination node.

- *F.A. Ahmed Jedidi is with CES Laboratory, National School of Engineering, University of Sfax, Tunisia*
- *S.B. Mohamed Abid is with CES Laboratory, National School of Engineering, University of Sfax, Tunisia.*

Based on multiplexing techniques that are being used to increase the transmission capacity in the optical fiber, AONs are divided into two types: OTDM and Wavelength-Division Multiplexed (WDM) networks. Specifically, Terahertz Optical Asymmetric Demultiplexer (TOAD) switch as an important component in OTDM systems has received wide interests in all-optical processing, switching and demultiplexing in OTDM networks. This is particularly because it has a simple structure, fast response time, and low switching energy [7]. However, TOADs are vulnerable to various forms of optical crosstalk. Then, they make a high degradation in original signal, exactly, because they are additive in nature.

In this paper, we propose a new method for detecting and localizing crosstalk in TOAD components. In Section II, we give a brief overview on optical crosstalk forms that may arise in TOAD switches. In Section III, we present the key concepts of Crosstalk Identification and Localization System (CILS). In Section IV, we propose the internal architecture and design of this system. Last, in Section V, we discuss the simulation results and present the open directions for future work.

## 2. CROSSTALK IN TOAD ROUTERS

Optical crosstalk is present in OTDM components and degrades the quality of signals, to increase their Bit Error Rates (BER) performance as they travel through the network. As a matter of fact, two forms of optical crosstalk can arise in TOADs node: interchannel crosstalk and intrachannel crosstalk [8]. Interchannel crosstalk occurs due to the appearance of non-target channel pulses in the switching window of the demultiplexer, and intrachannel crosstalk occurs due to transmission of energy from non-target channels during the carrier recovering period. For convenience, the centre of a switching window was defined at the middle between the two 50% maximum crossing points of the switching window for two windows TOAD. The peak of the target channel data pulse was then placed in the centre of the switching window Figure 1. For more details of the analysis of TAOD intrachannel and interchannel crosstalk can be found in [7].

A typical OTDM router is composed of two-terahertz optical asymmetric demultiplexers. Figure 2 shows the internal architecture of 2x2-TOAD parallel router. It consists of two 1x2 TOAD routers and two buffers. Optical buffers are used to eliminate contention at the outputs. Thus the transmitted data can be stored in the buffer or passed straight through without delay.

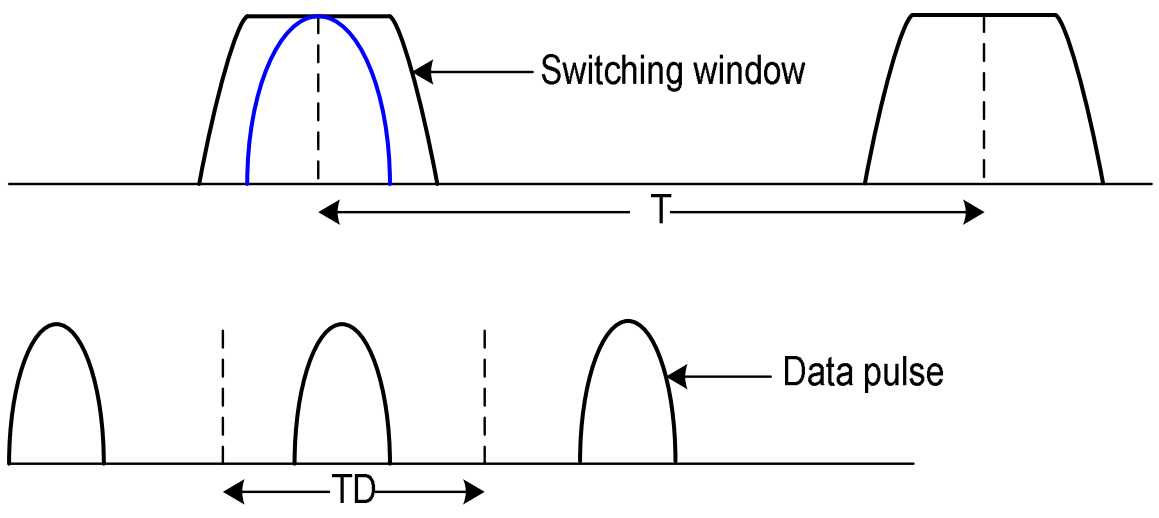

• Fig. 1: Switch window and data pulse

For example, when two optical packets arrive at the same time and must be routed simultaneously to the same output port, then only one packet can be switched to the output and the other is stored in the buffer [7]-[9].

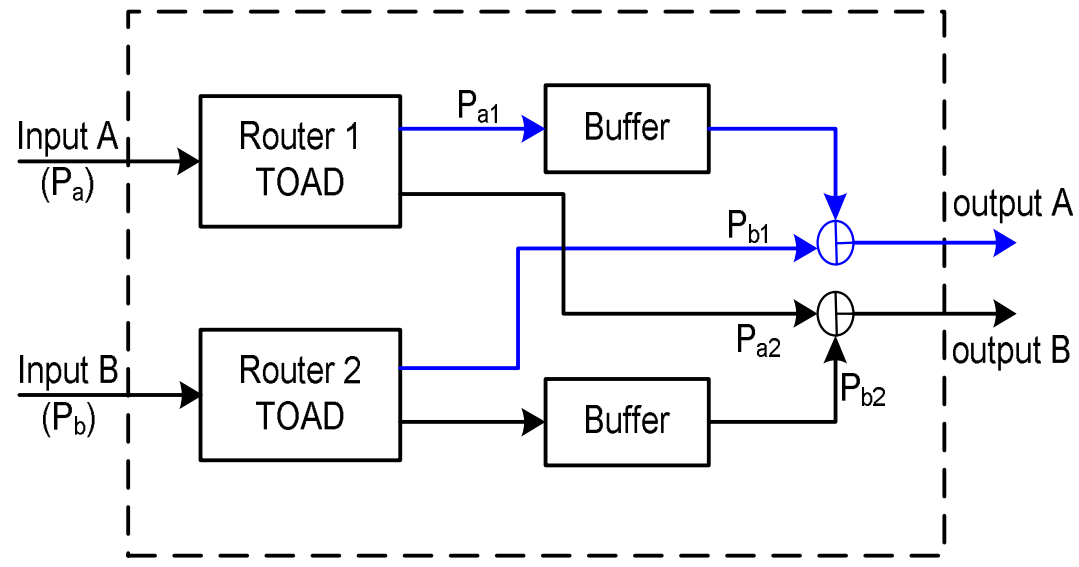

• Fig. 2: Parallel router TOAD architecture [7]

The signal power at the output of router1 is given by [9]:

$$P_{a1} = P_a \cdot (1 + X_{inter1} + X_{intra1}) \tag{1}$$

Where, $P_a$ is the input signal power, $X_{inter1}$ and $X_{intra1}$ are the interchannel and the intrachannel crosstalk components at the output of router1, respectively [9].

Similarly, the signal power at the output of router2 is given by: $P_{b1} = P_b \cdot (1 + X_{inter2} + X_{intra2})$. (2)

In this example it is assumed that packet from router1 is switched and the packet from router2 is stored in the buffer to avoid contention at the output A. Let Sa be the power of the signal [9]-[11].

$$S_a = \begin{cases} P_{a1} & \text{if packet form router1 is in the output A} \\ P_{a1} & \text{if packet form router2 is in the output A} \end{cases}$$

For simplification, it is assumed that the inputs powers $P_a$ and $P_b$ are the same and equal to $P_0$. It is also assumed that both crosstalk components $X_{inter}$ and $X_{intra}$ have the same levels at each router [9].

The total router crosstalk is then given by: $X_{tot} = X_{inter} + X_{intra}$. Then, the signal power at the output1 can be expressed by: $S_a = P_0 \cdot (1 + 2 \cdot X_{tot})$ (3)

Similarly, the output of n parallel routers can be is given by: $S_a = P_0 \cdot (1 + n \cdot X_{tot})$ (4)

The normalized crosstalk of parallel configuration is given in [9]:

$$X_n = \frac{S_a - P_0}{P_0} = n\,X_{tot} \quad (5)$$

## 3. CROSSTALK IDENTIFICATION AND LOCALIZATION SYSTEM

The main task of the CILS is to detect and localize optical crosstalk in TOAD routers. To identify the source and estimate the value of crosstalk, the CILS use routing information of channels on the input and output side of destination node of any established lightpath. With each passage of the optical signal of a TOAD router a crosstalk is added to it.

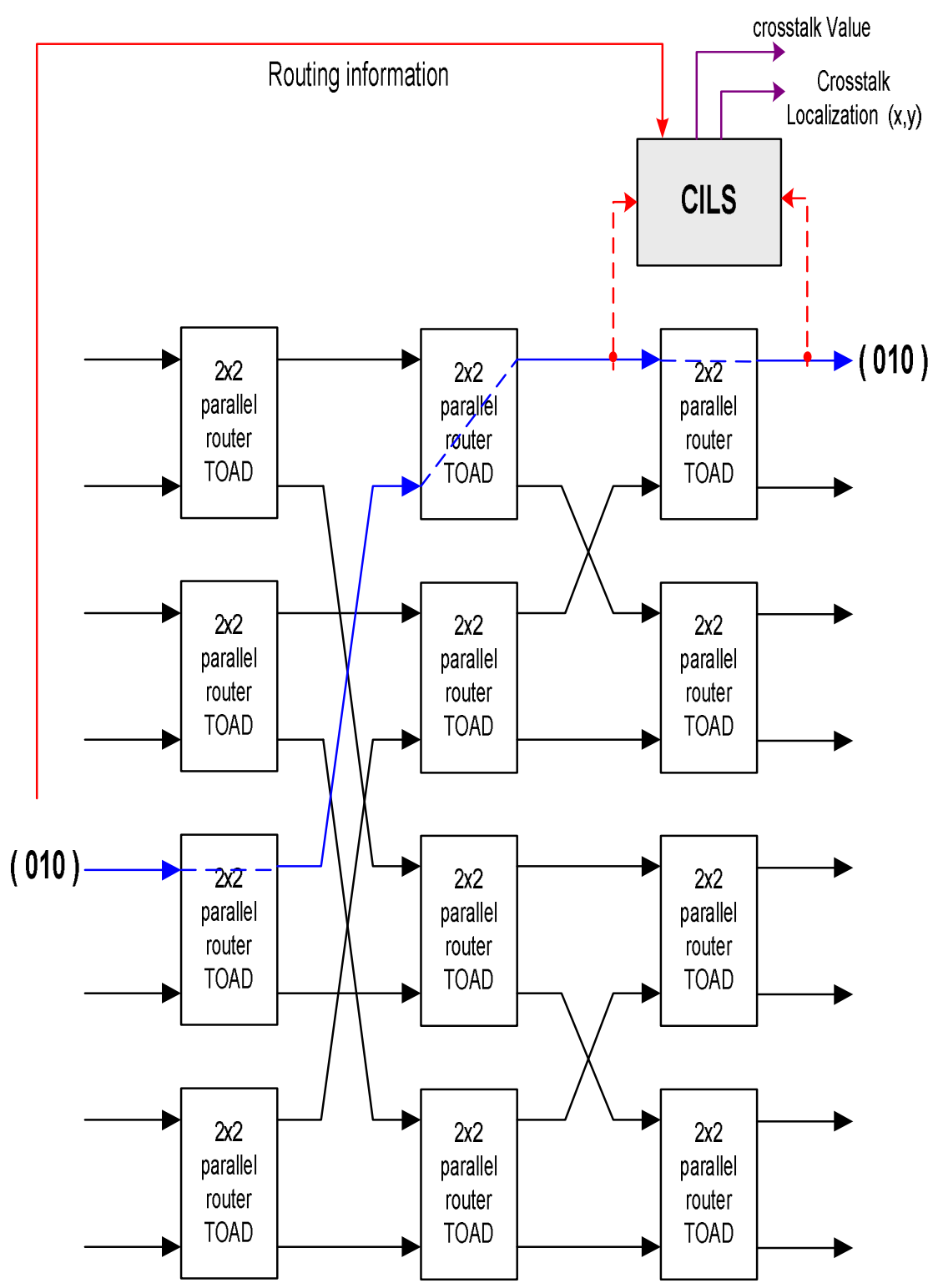

• Fig. 3: CILS in 8x8 Banyan network architecture

To understand quantitatively the principal functionality of CILS, we consider an 8x8 Banyan sample network which is composed of parallel routers as shows in Figure 3.

Before we explain our crosstalk identification and localization method we assume that every TOAD router in our network is basically a 2x2 TOAD parallel router.

If we consider the routing path (blue line in Figure 3, for example), we note in the worst case that at each passage from a stage to another value of crosstalk is added in signal. Consequently the value of the crosstalk depends on the number of stage and it is given in [9]:

$$X_n = \left(1 + X'_{tot}\right)^k - 1 \quad (6)$$

Where, k=1, 2, …,n represents the number of series stages and $X'_{tot}$ is the total router crosstalk. Then crosstalk can be expressed by: $X_n = (1 + n.X_{tot})^k - 1$ 7)

In our sample netwok, the 8x8 banyan network , k=3 and n=2: that is $X_n = (1 + 2.X_{tot})^3 - 1$. (8)

Figure 4 shows the flow chart of CILS which is composed of two main parts: the Crosstalk Identification Block and Crosstalk Localization Block.

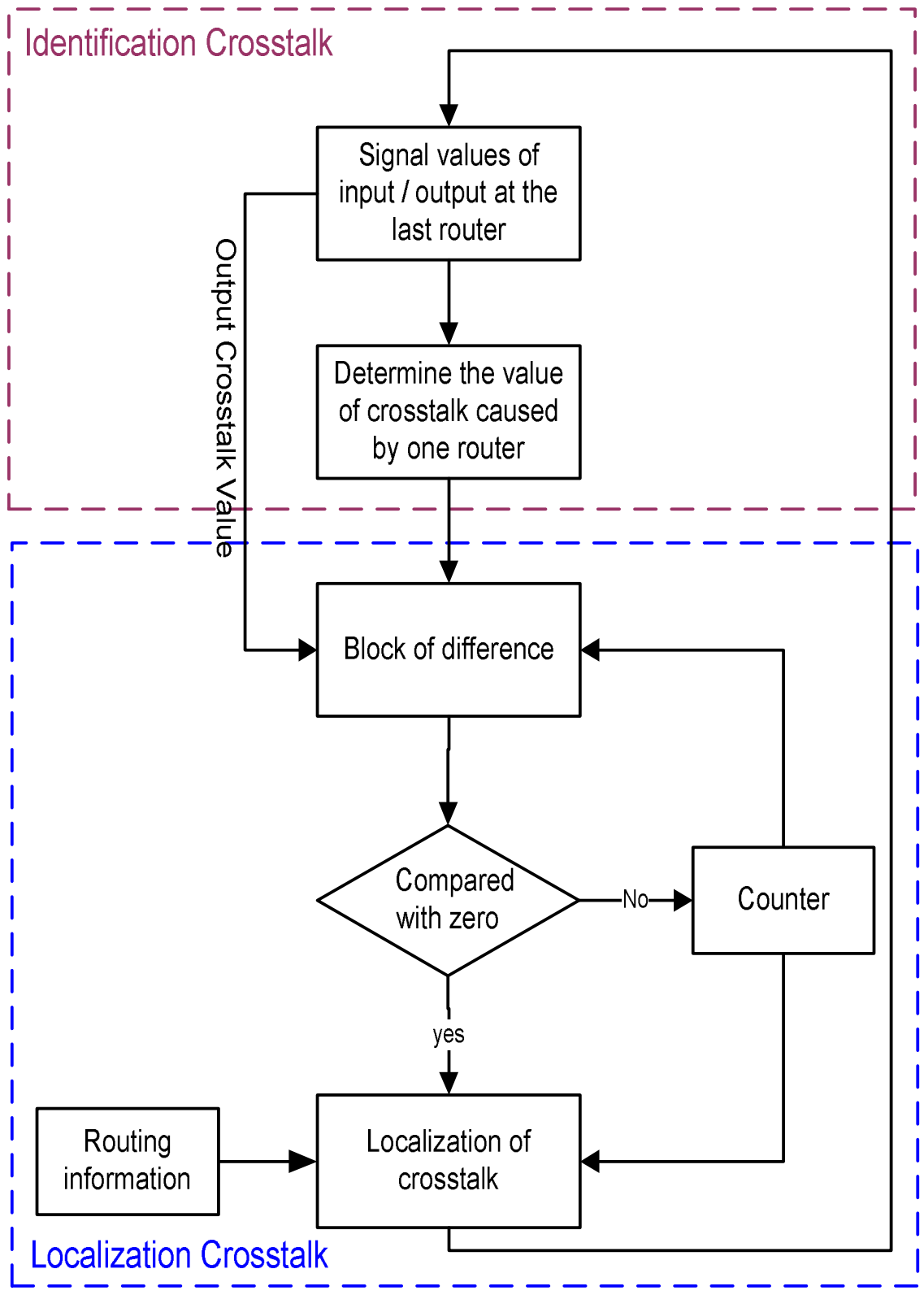

• *Fig. 4: Flow chart of CILS*

The Crosstalk Identification Block is responsible for the detection and determination of the crosstalk value. For that, we use the input output signal path passing through a destination node in a real-time fashion. We split off portions of signals for testing purpose. As shown in Figure 5, the taped optical signals are then photo-detected in the Optical Processing Block (OPB) and the resulting electrical signal is processed into Crosstalk Value Block (CVB). CVB calculates the value of crosstalk. Then, this mean value of crosstalk is used by the Crosstalk Localization part[10].

The Crosstalk Localization Block is composed by two processes. First, we make successive differences between the value of the crosstalk and the input signal up to zero. This block counts the number of different operation N. Second, we use this quantity and routing information to localize crosstalk in the network. In fact we follow the ligthpath described by the routing information and account the number of the nodes corresponding to N. Finally we localize the node that caused crosstalk precisely these coordinates (x,y). When x is the number of stage and y is the number of node in the correspondent stage.

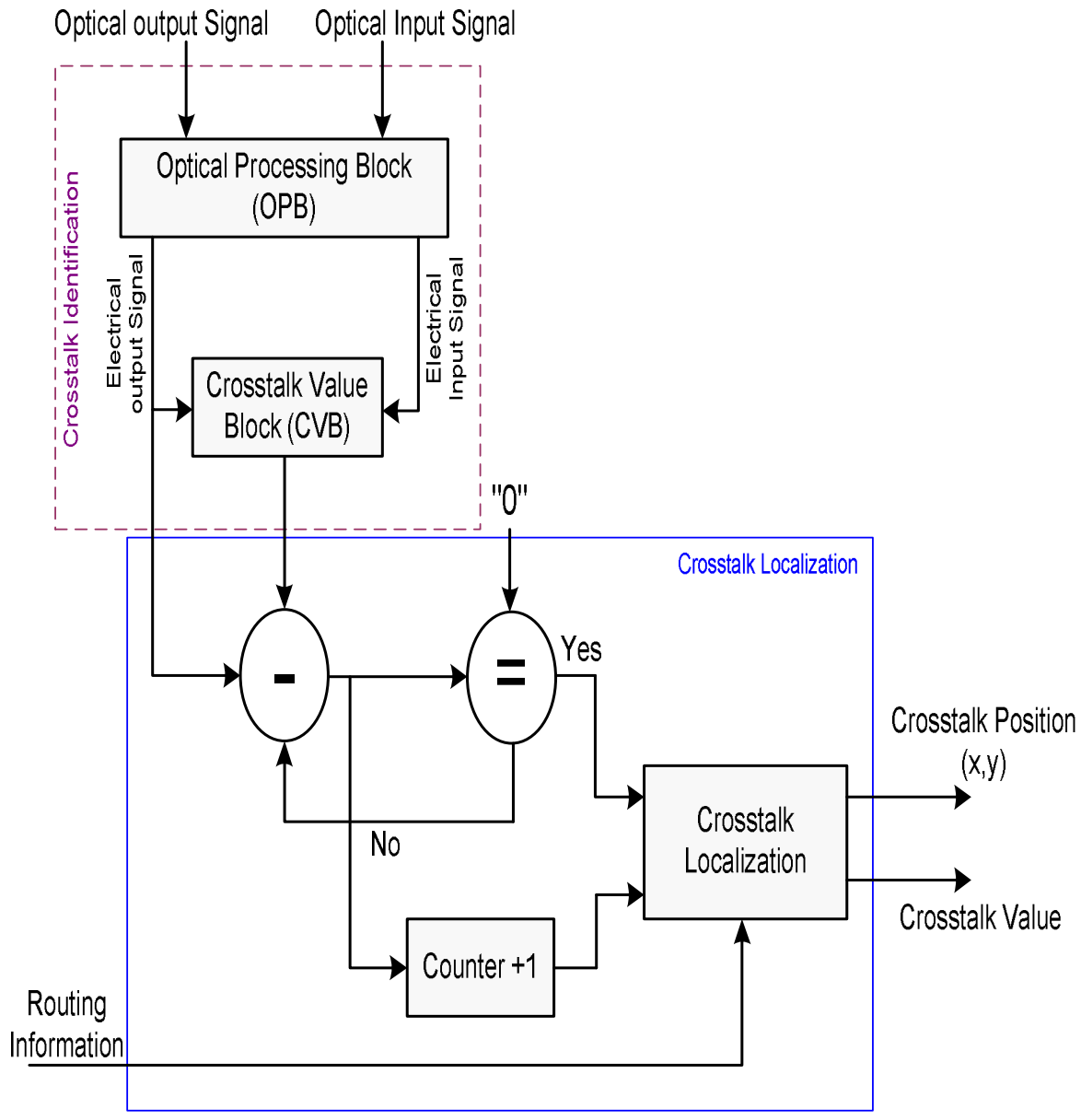

• Figure 5: Internal architecture of CILS

## 4. SIMULATION RESULTS

The performance evaluation of CILS is shown in Figure 6. The internal design and simulation of this device was performed by a hardware simulation tool (Project Navigator of Xilinx and ModelSim) with a frequency of 300 MHz. The red curve shows the execution times as a function of the number of stage in Banyan network in the worst case (that is crosstalk arises in each router stage). In this case, the execution time increases with the number of router stage. In 20x20 Banyan network case, the execution times is 74 ns therefore CILS is functional in real-time.

The blue curve shows the Lookup Table (LUT) as a function of the number of stage. LUT increase with the number of router stage because we have more and more operation blocks. Then, LUT reflects the occupation area of CILS in chip.

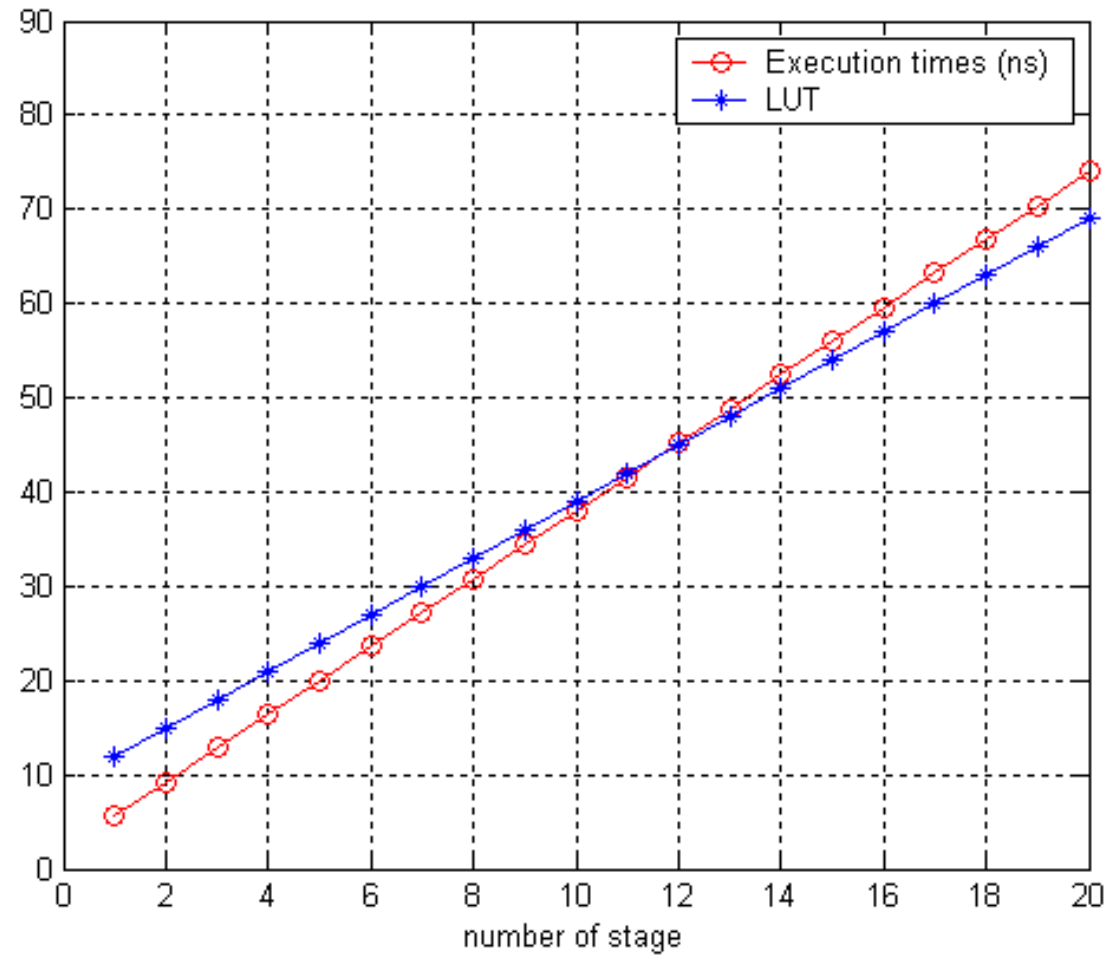

• Fig. 6: Execution times and LUT vs. number of stages for Banyan network

## 5. CONCLUSION

As more intelligence and control mechanisms are added to optical networks, the deployment of an efficient and secure management system, using suitable controlling and monitoring methods, is highly desirable. While some of the available management mechanisms are applicable to different types of network architectures, many of these are not adequate for AONs. An important implication of using AON components in communication systems is that available methods used to manage and monitor the health of the network may no longer be appropriate. Therefore, without additional control mechanisms a break in the core of an optical network might not be detectable.

In this paper we analyzed optical crosstalk forms that may arise in TOAD routers. Then, we proposed a new method that can be used for identifying and localizing crosstalk in TOAD components. This method can be used for supervising performance degradation in AON components offering the benefit of relaxing the high cost and complexity of signal quality monitoring for future AON management solutions.

**First A. Ahmed** Jedidi is currently a PhD student in Systems Engineering Informatics at the Nationale School of Engineers of Sfax , in the laboratory "Computer & Embedded Systems. He is currently working towards his Ph.D. in the detection and localization and estimation crosstalk in all-optical networks (AONs).
**Second B. Moahmed Abid**. is currently Professor at Sfax University in Tunisia. He holds a Diploma in Electrical Engineering in 1986 from the University of Sfax in Tunisia and received his Ph.D. degree in Computer Engineering in 1989 at University of Toulouse in France. His current research interests include Hardware-Software System on Chip co-design, reconfigurable FPGA, real time system and embedded system. He has authored/co-authored over 100 papers in international journals and conferences. He served on the technical program committees for several international conferences. He also served as a co-organizer of several international conferences.